\documentclass[a4paper,fleqn,usenatbib]{mnras}

\usepackage{newtxtext,newtxmath}

\usepackage[T1]{fontenc}
\usepackage{ae,aecompl}

\usepackage{graphicx}	
\usepackage{amsmath}	
\usepackage{amssymb}	
\usepackage{multicol}        
\usepackage{bm}		
\usepackage{pdflscape}	
\usepackage{url}

\title[Gaia DR2 Velocity Distribution]{Radial Distribution of Stellar Motions in {\it Gaia} DR2}

\author[D. Kawata et al.]{
Daisuke Kawata$^{1}$\thanks{E-mail: d.kawata@ucl.ca.uk}, Junichi Baba$^{2}$, Ioana Ciuc\u{a}$^{1, 6}$, Mark Cropper$^{1}$, 
\newauthor{Robert J. J. Grand$^{3,4}$, Jason A. S. Hunt$^{5}$, George Seabroke$^{1}$}
\\
$^{1}$Mullard Space Science Laboratory, University College London, Holmbury St. Mary, Dorking, Surrey, RH5 6NT, UK\\
$^{2}$National Astronomical Observatory of Japan, Mitaka, Tokyo 181-8588, Japan\\
$^{3}$Department of Astronomy and Astrophysics, University of Toronto, 50 St. George Street, Toronto, ON M5S 3H4, Canada\\
$^{4}$Department of Astronomy, The University of Tokyo, 7-3-1 Hongo, Bunkyo-ku, Tokyo 113-0033, Japan\\
$^{5}$Dunlap Institute for Astronomy and Astrophysics, University of Toronto, Ontario M5S 3H4, Canada\\
$^{6}$ LSSTC Data Science Fellow\\
}

\date{Accepted XXX. Received YYY; in original form ZZZ}

\pubyear{2018}

\begin{document}
\label{firstpage}
\pagerange{\pageref{firstpage}--\pageref{lastpage}}
\maketitle
%
\begin{abstract}
By taking advantage of the superb measurements of position and velocity for an unprecedented large number of stars provided in {\it Gaia} DR2, we have generated the first maps of the rotation velocity, $V_{\rm rot}$, and vertical velocity, $V_{\rm z}$, distributions as a function of the Galactocentric radius, $R_{\rm gal}$, across a radial range of $5<R_{\rm gal}<12$~kpc. In the $R-V_{\rm rot}$ map, we have identified many diagonal ridge features, which are compared with the location of the spiral arms and the expected outer Lindblad resonance of the Galactic bar. We have detected also radial wave-like oscillations of the peak of the vertical velocity distribution.
\end{abstract}




\section{Introduction}
\label{sec:intro}

Stellar velocity structure as a function of the Galactocentric radius, $R_{\rm gal}$, and the azimuthal position of the disc provides fruitful information about the impact of non-axisymmetric structures, such as the bar and spiral arms \citep[e.g.][]{wd00,khgpc14,Monari+16} and satellite galaxies \citep[e.g.][]{Gomez+12,DOnghia+16} on the Galactic disc. Recent ground-based spectroscopic surveys of Galactic stars with multi-object spectrographs have demonstrated the complex structure of stellar velocity fields including the velocity fluctuation of the Galactic disc \citep[e.g.][]{Widrow+12,bbgmnz15,Tian+17}, asymmetric motions \citep[e.g.][]{Wang+18,Williams+13,Carrillo+18} and resonance features \citep[e.g.][]{Liu+12,GMO13,Tian+17}. However, these studies are mainly based on the line-of-sight radial velocity only. Furthermore, the distance measurements rely on the photometric distance which are subject to dust extinction corrections. 

The European Space Agency's {\it Gaia} mission \citep{Gaia+Prusti16} has made their second data release \citep[{\it Gaia} DR2;][]{Gaia+Brown+18} which provides the unprecedentedly accurate measurements of parallax and proper motion \citep{Lindegren+18} and line-of-sight velocity of a large number of bright stars \citep{Cropper+18,Katz+RV+18,Sartoretti+18}. This revolutionary data set provides six dimensional phase space information: the positions and velocities of stars, which allows us to measure the Galactic rotation, radial and vertical velocity structure in different regions of the Galactic disc, as demonstrated in \citet{Gaia+Katz18Disc}. The line-of-sight velocities are available only for the bright ($G < \sim13$~mag) stars in the {\it Gaia} DR2. However, for fainter stars accurate parallax and proper motions are still available. As demonstrated in \citet{Hunt+17} with the {\it Gaia} DR1, we can use the proper motion of the Galactic longitudinal direction, $V_{\rm l}$, as a proxy to the Galactic rotation velocity, $V_{\rm rot}$, in the direction of $l=0$ and 180~deg and $b=0$.  \citet{Hunt+17} identified a fast rotating moving group which spreads over 0.6~kpc in radius, and demonstrated that the decreasing rotation velocity of the {\it Hercules} stream found in \citet{Monari+17} can be seen in $R_{\rm gal}$ vs. $V_{\rm l}$ without the line-of-sight velocity information. \citet{Schoenrich+Dehnen18} further analysed both $V_{\rm l}$ and velocity in the direction of the Galactic latitude, $V_{\rm b}$, as a proxy of vertical motion, $V_{\rm z}$, using the {\it Gaia} DR1 in the direction of $l=0$ and $l=180$~deg. They found that $\langle V_{\rm z} \rangle$ as a function of the guiding centre exhibits wave-like oscillations superposed on top of an overall increase with radius, which they interpret as a combination of vertical waves propagating radially and the warp of the disc. 

In this {\it Letter}, we take advantage of the superb astrometric accuracy of {\it Gaia} DR2 to produce the first maps of $V_{\rm rot}(\sim V_{\rm l})$ and $V_{\rm z}(\sim V_{\rm b})$ distributions as a function of Galactocentric radius, covering the radial range $5 \lesssim R_{\rm gal} \lesssim 12$~kpc, in the direction of $l=0$ and $l=180$ and $b=0$. We identify diagonal ridge features in $R-V_{\rm rot}$ map and wave-like features in the $R-V_{\rm z}$ map, and compare them with the location of the spiral arm and the resonance radii of the expected bar pattern speed. 
 
 Section~\ref{sec:data} describes our data and sample selection. Section~\ref{sec:res} shows our results. A summary and discussion of this study are presented in Section~\ref{sec:sum}.

\section{Data and Analysis}
\label{sec:data}

We have extracted two different samples of stars from the {\it Gaia} DR2 catalogue in a volume within the width in the disc plane of $0.2$~kpc and height from the plane within $0.2$~kpc along the line of the Galactic centre and the Galactic anti-centre. We assumed the Sun's Galactocentric radius of $R_0=8.2$~kpc and vertical offset from the Galactic mid plane of $z_0=25$~pc. We also assumed the solar motion in the rotation direction of $V_{\rm \sun}=248$~km~s$^{-1}$ which is faster than the rotation speed of the Local Standard of the Rest (LSR), $V_{\rm LSR}$ by $V_{\rm \sun}=11$~km~s$^{-1}$. We also assumed the solar motion in the vertical direction of $W_{\rm \sun}=7.0$~km~s$^{-1}$. These values are taken from \citet{bhg16}. Note that these values are set for convenience of presentation, and our results do not depend on these assumed values. 
 
The first sample comprises stars whose line-of-sight velocity measured with the {\it Gaia}'s RVS instrument \citep{Cropper+18} are available in the {\it Gaia} DR2. We also selected stars whose radial velocity uncertainties are smaller than 5~km~s$^{-1}$ and whose parallax accuracy is better than 15~\%, i.e. ${\varpi}/\sigma_{\varpi}>1/0.15$, where $\varpi$ is parallax and $\sigma_{\varpi}$ is its uncertainty. As mentioned above, we only selected stars within 0.2~kpc from the plane and 0.2 kpc perpendicular to the line of the Galactic centre and the Galactic anti-centre. For this sample, the full six dimensional position and velocity information is available, and $V_{\rm rot}$ and $V_{\rm z}$ are derived with the assumed Galactic parameters shown above. We used {\tt galpy} \citep{jb15} for all coordinate transformation. There are 861,680 stars in this sample. We call this sample the ``RVS'' sample. 

The second sample includes all stars brighter than $G=15.2$~mag, but again with ${\varpi}/\sigma_{\varpi}>1/0.15$. This sample has no line-of-sight velocity information in the Gaia DR2, except the bright stars with limited effective temperature values. Hence, we limit the sample to within $|b|<10$~deg and $|l|<10$~deg or $|l|-180<10$~deg. Again, we only selected stars within 0.2~kpc from the plane and 0.2 kpc from the line of the Galactic centre and the Galactic anti-centre. In this limited region, there are 1,049,340 stars. We call this sample the ``All'' sample. In this {\it Letter}, we assume that in this limited angular region $V_{\rm l}=V_{\rm rot}$ and $V_{\rm b}=V_{\rm z}$. 
 
 Using the mock data constructed with {\tt Galaxia} \citep{sbhjb11}, we estimated that the average difference between $V_{\rm l}$ and $V_{\rm rot}$ is about 0.3~km~s$^{-1}$. However, the average differences depend on $l$ and it increases to about 2.7~km~s$^{-1}$ at $|l|=10$~deg or $|l-180|=10$~deg. Because we discuss the $V_{\rm rot}$ distribution as a function of $R_{\rm gal}$ by summing the contribution from all the stars with different $l$, this systematic dependence on $l$ should not affect our results. The average difference between $V_{\rm b}$ and $V_{\rm z}$ is smaller than 0.4~km~s$^{-1}$, and we do not find any correlation with $l$. This is consistent with what is shown in \citet{Schoenrich+Dehnen18}. 
 
\begin{figure*}
 \includegraphics[width=\hsize]{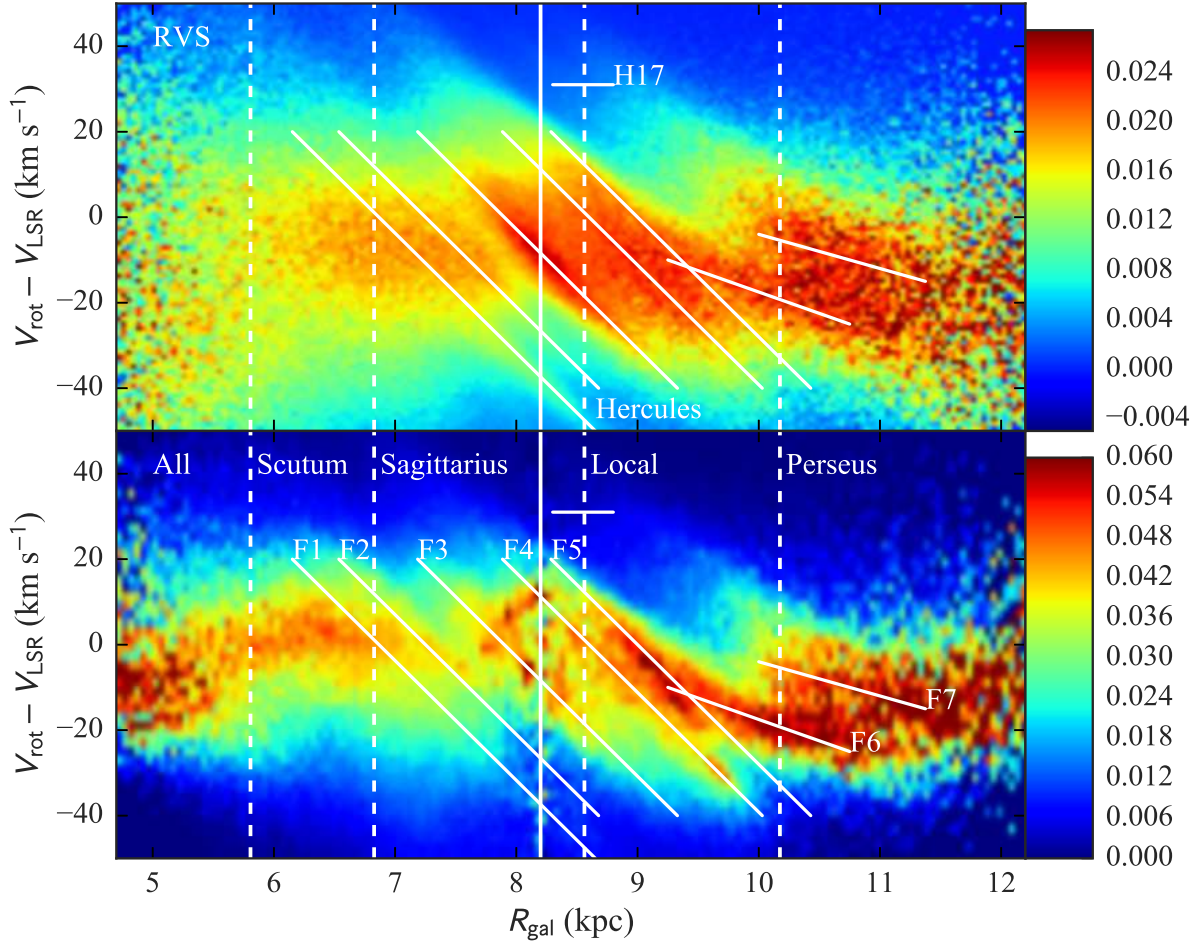}
\caption{Normalised distribution of the rotation velocity for our RVS stars (upper) and our All stars (lower) as a function of the Galactocentric radius. The vertical dashed lines show the position of the Scutum, Sagittarius, Local and Perseus spiral arms from left to right, which are calculated from \citet{rmbzd14}. The vertical solid line is the assumed solar radius. The white lines highlight the identified ridge features.}
\label{fig:rvrot}
\end{figure*}

\section{Results}
\label{sec:res}

\subsection{$R_{\rm gal}$ vs.\ $V_{\rm rot}$}
\label{sec:rVrot}

Fig.~\ref{fig:rvrot} shows the distribution of $V_{\rm rot}-V_{\rm LSR}$ as a function of $R_{\rm gal}$ for our two samples of stars. Because each sample has a different number of stars distributed in different radial bins, we normalised the distribution at each radial bin to highlight the features in the velocity distribution. Because the brightness limits are different for each sample, the radial range covered by each sample is different; stars in our ``All'' sample reach up to $\sim 4$ kpc from the Sun, whereas stars in our ``RVS'' sample are confined to a slightly smaller volume. It is striking to see many diagonal ridge-like features, highlighted by white diagonal lines.
To our knowledge, this is the first time that these clear features are seen in observational data, and this is a new window opened up by {\it Gaia}. Features are more clear in our ``All'' sample (except F1, F2 and F3 which are more clear and selected in the ``RVS'' sample around the solar radius), and are therefore selected by eye in the ``All'' sample panel. The ``All'' sample includes fainter stars and there are more stars closer to the disc mid plane. Therefore, more features are visible in this sample. F1 and F2 correspond to the split Hercules streams \citep[see also][]{Gaia+Katz18Disc,Antoja+18,Trick+18}, which are highlighted in the upper panel. F3 is due to the so-called Hyades and Pleiades moving groups, and F4 corresponds to the Sirius moving group \citep[see also][]{Ramos+18}. F2, F3, F4 and F5 are highlighted between $V_{\rm rot}-V_{\rm LSR}=20$ and $-40$~km~s$^{-1}$, because most of the features cover this velocity range, although some features extend to higher velocity. F1 extends to lower velocities, which is clearly visible in the ``RVS'' sample. F6 and F7 are highlighted only in the range visible in the ``All'' sample. 

The vertical dashed lines in Fig.~\ref{fig:rvrot} show the position of the spiral arms at $y=0$. The positions are calculated from what is measured in \citet{rmbzd14} and scaled to our assumed $R_0=8.2$~kpc. Each line corresponds to the Scutum, Sagittarius, Local and Perseus spiral arms from left to right. We can see the two split inclined features with 
systematically higher and lower rotation velocity than the LSR rotation speed at the location of Perseus arms (F6 and F7). Although it
is tentative, 
especially in the ``All'' sample the rotation velocity in the inside of the Scutum arm
is clearly slower than that in the outside of the arm.
These bimodal features either side of the LSR rotation speed are expected around the spiral arm at the co-rotation resonance \citep[e.g.][]{khgpc14}. 
If these are from co-rotation of the spiral arms, this indicates that the Scutum and Perseus arms have different pattern speeds, which can be naturally explained if these spiral arms are co-rotating at every radius, as seen in recent $N$-body simulations \citep{wbs11,gkc12a,gkc12b,bsw13}. 
However, to test the spiral arm scenario, we need to look at $V_{\rm rot}$ distribution in a larger region of the disc \citep{hkgmpc15,Quillen+18,Hunt+Hong+Bovy+18}. 

There is similar bimodal feature around the Local arm (F3 and F4). The slope of this feature ($dV_{\rm rot}/dR\sim-28$~km~s$^{-1}$~kpc$^{-1}$) looks steeper than the one in the Perseus arms ($dV_{\rm rot}/dR\sim-9$~km~s$^{-1}$~kpc$^{-1}$). The Local arm is often considered to be a weak spiral arm or spur, and it is not expected to influence the stellar motion as strongly as the main spiral arms, like the Scutum and Perseus arms where clear stellar density enhancements are observed. Hence, we expect that the origins of F3 and F4 are not related to the spiral arms \citep[but see][for an alternative view explaining these features with the Local arm and the Local spur]{Quillen+18}.


Compared to the Scutum and Perseus arms, there is no such feature at the radius of the Sagittarius arm, except the extension of F1 and F2. Therefore, we speculate that these arms are not stellar arms, but only gaseous star-forming arms, as indicated in \citet{bcbim05}, who found no significant density enhancement at the position of the Sagittarius arm in the Spitzer GLIMPSE survey. In this case, the spiral arms may not have enough gravitational potential to influence the stellar motions. If this is true, our results support $m=2$ spiral arms in the Milky Way, which is in fact more common in a barred galaxy \citep[e.g.][]{Hart+17}.

We also note that we can see also a group of stars (indicated as ``H17'' in the upper panel of Fig.~\ref{fig:rvrot}) with high rotation velocities just outside of $R_0$, which were found in \citet{Hunt+17}. However, they are not a horizontal feature as suggested in \citet{Hunt+17}, but rather form a diagonal feature parallel to F5 with a higher rotation speed in the ``RVS'' sample. This feature is tentative, therefore we do not select it as a clear diagonal feature. We note that this feature seems to be connected to the ``Arch 1'' feature in the solar neighbourhood velocity distribution, highlighted in the wavelet analysis in \citet{Ramos+18}, although they do not find any extension of the feature to regions outside of the solar radius. Nevertheless, it will be interesting to study this feature further using future {\it Gaia} data releases.

Fig.~\ref{fig:rvrot} also traces the whole resonance feature of the Hercules stream. With the {\it Gaia} DR1 and LAMOST data, \citet{Monari+17} found the rotation speed of the Hercules stream ($V_{\rm rot}-V_{\rm LSR}\sim -30 {\rm\  to}-40$~km~s$^{-1}$ at $R=R_0$) decreases with radius for $R_{\rm gal}>R_{\rm 0}$, and therefore the gap between the Hercules stream and the Hyades and Pleiades moving groups (F3 in Fig.~\ref{fig:rvrot}) decreases with radius. It is expected that the gap should extend to the inner disk \citep{Antoja+14} and that the rotation speed of the Hercules stream increases with decreasing $R_{\rm gal}$. For the first time, the {\it Gaia} DR2 has revealed the inner extension of the gap due to the Hercules stream as a clear gap between F2 and F3 in Fig.~\ref{fig:rvrot}. This gap crosses the $V_{\rm LSR}$ at $R_{\rm gal}\sim7.6$~kpc. 
This could be the OLR of the fast rotating bar, as widely believed \citep{wd00,Monari+17}. 
However, we note that there are many mechanisms which can explain the Hercules stream feature \citep[e.g.][]{Hattori+18,Hunt+Bovy18}, and examining the velocity distribution in a larger region of the disc is necessary to determine the pattern speed of the bar and the location of its resonances.

\begin{figure}
 \includegraphics[width=\hsize]{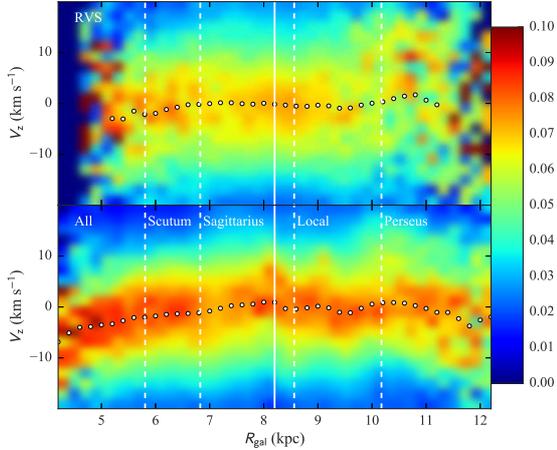}
\caption{Distribution of the vertical velocity for the ``RVS'' stars (upper) and stars in the ``All'' sample (lower) as a function of Galactocentric radius. The symbols indicate the peak of the density at each radius. The circles show the stronger peak detected with 2 Gaussian models. The vertical dashed lines show the position of the Scutum, Sagittarius, Local and Perseus spiral arms from left to right, as suggested in \citet{rmbzd14}.  The vertical solid line indicates the solar radius.}
\label{fig:rvz}
\end{figure}

\begin{figure}
 \includegraphics[width=\hsize]{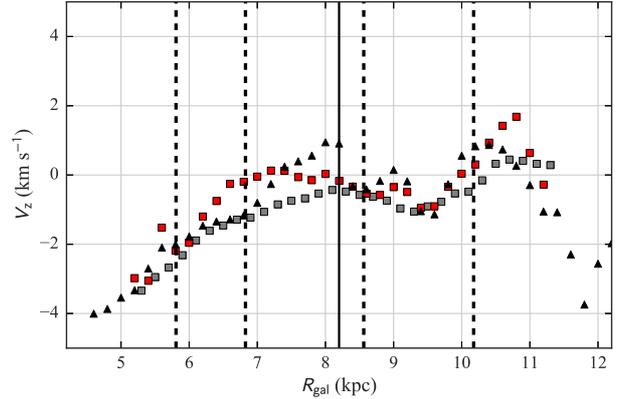}
\caption{The mean vertical velocity of the most significantly identified feature in Fig.~\ref{fig:rvz}. Red squares and black triangles are the results for the ``RVS'' sample and the ``All'' sample, respectively. For a reference, grey squares show the median $V_{\rm z}$ for the ``RVS'' sample.}
\label{fig:rvzpeak}
\end{figure}
 
\subsection{$R_{\rm gal}$ vs.\ $V_{\rm z}$}
\label{sec:rVz}

Fig.~\ref{fig:rvz} shows the distribution of $V_{\rm z}$ as a function of $R_{\rm gal}$ for our three samples of stars. We again normalised the distribution at each radial bin to highlight the features in the distribution. We do not find any features like the ridge features in $V_{\rm rot}$. Instead, the mean velocity shows wave-like oscillations and increases with $R_{\rm gal}$. To trace the centroid of the velocity distribution, we selected stars within 0.2 kpc from radial grid points at every 0.2 kpc in $R_{\rm gal}$, i.e. each grid point has 0.1 kpc of overlap region with their neighbour points, and hence every second grid point is independent. At each radial bin, we fit the velocity distribution with two Gaussians using extreme-deconvolution \citep{Bovy+11}. 
We estimated the uncertainty of $V_{\rm z}$ (or $V_{\rm b}$) by taking 1,000 Monte-Carlo (MC) samples of the parallax and proper motion with their uncertainties and correlations for each star, converting them to $V_{\rm z}$ (or $V_{\rm b}$) and taking the standard deviation. Using different numbers of Gaussian models, we find that two Gaussians is the optimal and most robust choice to trace the peak of the velocity distribution. Fig.~\ref{fig:rvzpeak} shows that the position of the mean of the main Gaussian model, which show a clear oscillatory pattern \citep[see also Fig.~14][who showed similar results for the ``RVS'' sample in more detail]{Gaia+Katz18Disc}.
Interestingly, the results for the ``All'' sample are similar to what is seen in ``basic method'' of Fig.~9 in \citet{Schoenrich+Dehnen18}.  The peaks of the oscillatory pattern in Fig.~\ref{fig:rvzpeak} are around $R_{\rm gal}=8$ and 10.5~kpc, and the dip is around $R_{\rm gal}=9$~kpc with a small spike, which are similar in locations of the peak, dip and spike found in \citet{Schoenrich+Dehnen18}, although they used the guiding centre to detect these trends from the local sample of the {\it Gaia} DR1.  
This oscillatory pattern is superposed on top of a clear increase in $V_{\rm z}$ with $R_{\rm gal}$, which is suggestive of a warp in the outer disc as discussed in \citet{Schoenrich+Dehnen18}, and consistent with the predicted vertical waves induced by the Sagittarius dwarf galaxy in \citet{GMO13}.

Interestingly, our ``RVS'' sample shows a similar oscillatory pattern, but it has a slightly longer wavelength. As mentioned above, the ``All'' sample contains more stars closer to the disc mid plane compared to the ``RVS'' sample. We wildly speculate that this may indicate that stars in the denser plane have shorter wavelength than the stars above the plane, or that there may be multiple modes of the waves propagating differently for different stellar populations. 

Note that grey squares in Fig.~\ref{fig:rvzpeak} show the median $V_{\rm z}$ for the ``RVS'' sample as a comparison to the mean of the main Gaussian model shown with red squares. The median $V_{\rm z}$ also shows the oscillatory pattern. However, the amplitude is smaller and the spike at $R_{\rm gal}=9$~kpc is not seen. Hence, we think that deconvolution of the velocity distribution is important to get rid of the kinematically hot component, like halo and thick disc stars, and highlight the main velocity features in the thin disc.

\section{Summary}
\label{sec:sum}

By taking advantage of the unprecedented precise astrometric measurements of a large number of stars provided by the {\it Gaia} DR2, we have generated the first maps of the $R-V_{\rm rot}$ and $R-V_{\rm z}$ covering a radial range of 5 to 12 kpc in Galactocentric radius along the Galactic centre and Galactic anticentre line-of-sight. We discovered many diagonal ridge features in the $R-V_{\rm rot}$ map. Some of these are likely related to the perturbations from the bar's outer Lindblad resonance (OLR) and spiral arm. Alternatively, as suggested in \citet{Antoja+18}, some of these features could be due to phase-wrapping \citep{Minchev+09,Gomez+12}.  We found the transition of $V_{\rm rot}$ between the inside and the outside of the Scutum and Perseus arms. We speculate that these features are due to co-rotation resonances of the spiral arms, which may be explained with the transient spiral arm scenario. There are several ridge features around the solar neighbourhood/Local arm, but the features are steeper compared to the ridges around the Perseus arm. We speculate that ridges of different slopes have different origins, and more theoretical works are required to explain these features.

In the $R-V_{\rm z}$ distribution, we found the peak of the $V_{\rm z}$ distribution shows wave-like features almost identical to those seen in the local sample of {\it Gaia} DR1 in \citet{Schoenrich+Dehnen18}. The origin of the wave modes must be tightly related to the formation and evolution of the Galaxy \citep[e.g.][among others]{Widrow+12,GMO13,delaVega+15,XNC15,GWG16}, and comparisons between these observations and models are urgently required.

 \section*{Acknowledgments}
We thank an anonymous referee for their constructive comments and helpful suggestions which have improved the manuscript.
DK, IC, MC and GS acknowledge the support of the UK's Science \& Technology Facilities Council (STFC Grant ST/N000811/1). JB is supported by the Japan Society for the Promotion of Science (JSPS) Grant-in-Aid for Scientific Research (C) Grant Number 18K03711. IC is also grateful the STFC Doctoral Training Partnerships Grant (ST/N504488/1). RJJG acknowledges support by the DFG Research Centre SFB-881 'The Milky Way System', through project A1. JH is supported by a Dunlap Fellowship at the Dunlap Institute for Astronomy \& Astrophysics, funded through an endowment established by the Dunlap family and the University of Toronto. This work was inspired from our numerical simulation studies used the UCL facility Grace and the DiRAC Data Analytic system at the University of Cambridge, operated by the University of Cambridge High Performance Computing Service on behalf of the STFC DiRAC HPC Facility (\url{www.dirac.ac.uk}). This equipment was funded by BIS National E-infrastructure capital grant (ST/K001590/1), STFC capital grants ST/H008861/1 and ST/H00887X/1, and STFC DiRAC Operations grant ST/K00333X/1. DiRAC is part of the National E-Infrastructure. This work has made use of data from the European Space Agency (ESA) mission {\it Gaia} (\url{https://www.cosmos.esa.int/gaia}), processed by the {\it Gaia} Data Processing and Analysis Consortium (DPAC, \url{https://www.cosmos.esa.int/web/gaia/dpac/consortium}). Funding for the DPAC has been provided by national institutions, in particular the institutions participating in the {\it Gaia} Multilateral Agreement.   



\bibliographystyle{mnras}
\bibliography{./dkref} 

\begin{thebibliography}{}
\makeatletter
\relax
\def\mn@urlcharsother{\let\do\@makeother \do\$\do\&\do\#\do\^\do\_\do\%\do\~}
\def\mn@doi{\begingroup\mn@urlcharsother \@ifnextchar [ {\mn@doi@}
  {\mn@doi@[]}}
\def\mn@doi@[#1]#2{\def\@tempa{#1}\ifx\@tempa\@empty \href
  {http://dx.doi.org/#2} {doi:#2}\else \href {http://dx.doi.org/#2} {#1}\fi
  \endgroup}
\def\mn@eprint#1#2{\mn@eprint@#1:#2::\@nil}
\def\mn@eprint@arXiv#1{\href {http://arxiv.org/abs/#1} {{\tt arXiv:#1}}}
\def\mn@eprint@dblp#1{\href {http://dblp.uni-trier.de/rec/bibtex/#1.xml}
  {dblp:#1}}
\def\mn@eprint@#1:#2:#3:#4\@nil{\def\@tempa {#1}\def\@tempb {#2}\def\@tempc
  {#3}\ifx \@tempc \@empty \let \@tempc \@tempb \let \@tempb \@tempa \fi \ifx
  \@tempb \@empty \def\@tempb {arXiv}\fi \@ifundefined
  {mn@eprint@\@tempb}{\@tempb:\@tempc}{\expandafter \expandafter \csname
  mn@eprint@\@tempb\endcsname \expandafter{\@tempc}}}

\bibitem[\protect\citeauthoryear{{Antoja} et~al.,}{{Antoja}
  et~al.}{2014}]{Antoja+14}
{Antoja} T.,  et~al., 2014, \mn@doi [\aap] {10.1051/0004-6361/201322623}, \href
  {http://adsabs.harvard.edu/abs/2014A%26A...563A..60A} {563, A60}

\bibitem[\protect\citeauthoryear{{Antoja} et~al.,}{{Antoja}
  et~al.}{2018}]{Antoja+18}
{Antoja} T.,  et~al., 2018, preprint, \href
  {http://adsabs.harvard.edu/abs/2018arXiv180410196A} {} (\mn@eprint {arXiv}
  {1804.10196})

\bibitem[\protect\citeauthoryear{{Baba}, {Saitoh}  \& {Wada}}{{Baba}
  et~al.}{2013}]{bsw13}
{Baba} J.,  {Saitoh} T.~R.,   {Wada} K.,  2013, \mn@doi [\apj]
  {10.1088/0004-637X/763/1/46}, \href
  {http://adsabs.harvard.edu/abs/2013ApJ...763...46B} {763, 46}

\bibitem[\protect\citeauthoryear{{Benjamin} et~al.,}{{Benjamin}
  et~al.}{2005}]{bcbim05}
{Benjamin} R.~A.,  et~al., 2005, \mn@doi [\apjl] {10.1086/491785}, \href
  {http://adsabs.harvard.edu/abs/2005ApJ...630L.149B} {630, L149}

\bibitem[\protect\citeauthoryear{{Bland-Hawthorn} \&
  {Gerhard}}{{Bland-Hawthorn} \& {Gerhard}}{2016}]{bhg16}
{Bland-Hawthorn} J.,  {Gerhard} O.,  2016, \mn@doi [\araa]
  {10.1146/annurev-astro-081915-023441}, \href
  {http://adsabs.harvard.edu/abs/2016ARA%26A..54..529B} {54, 529}

\bibitem[\protect\citeauthoryear{{Bovy}}{{Bovy}}{2015}]{jb15}
{Bovy} J.,  2015, \mn@doi [\apjs] {10.1088/0067-0049/216/2/29}, \href
  {http://adsabs.harvard.edu/abs/2015ApJS..216...29B} {216, 29}

\bibitem[\protect\citeauthoryear{{Bovy}, {Hogg}  \& {Roweis}}{{Bovy}
  et~al.}{2011}]{Bovy+11}
{Bovy} J.,  {Hogg} D.~W.,   {Roweis} S.~T.,  2011, \mn@doi [Annals of Applied
  Statistics] {10.1214/10-AOAS439}, \href
  {http://adsabs.harvard.edu/abs/2011AnApS...5.1657B} {5}

\bibitem[\protect\citeauthoryear{{Bovy}, {Bird}, {Garc{\'{\i}}a P{\'e}rez},
  {Majewski}, {Nidever}  \& {Zasowski}}{{Bovy} et~al.}{2015}]{bbgmnz15}
{Bovy} J.,  {Bird} J.~C.,  {Garc{\'{\i}}a P{\'e}rez} A.~E.,  {Majewski} S.~R.,
  {Nidever} D.~L.,   {Zasowski} G.,  2015, \mn@doi [\apj]
  {10.1088/0004-637X/800/2/83}, \href
  {http://adsabs.harvard.edu/abs/2015ApJ...800...83B} {800, 83}

\bibitem[\protect\citeauthoryear{{Carrillo} et~al.,}{{Carrillo}
  et~al.}{2018}]{Carrillo+18}
{Carrillo} I.,  et~al., 2018, \mn@doi [\mnras] {10.1093/mnras/stx3342}, \href
  {http://adsabs.harvard.edu/abs/2018MNRAS.475.2679C} {475, 2679}

\bibitem[\protect\citeauthoryear{{Cropper} et~al.,}{{Cropper}
  et~al.}{2018}]{Cropper+18}
{Cropper} M.,  et~al., 2018, preprint, \href
  {http://adsabs.harvard.edu/abs/2018arXiv180409369C} {} (\mn@eprint {arXiv}
  {1804.09369})

\bibitem[\protect\citeauthoryear{{D'Onghia}, {Madau}, {Vera-Ciro}, {Quillen}
  \& {Hernquist}}{{D'Onghia} et~al.}{2016}]{DOnghia+16}
{D'Onghia} E.,  {Madau} P.,  {Vera-Ciro} C.,  {Quillen} A.,   {Hernquist} L.,
  2016, \mn@doi [\apj] {10.3847/0004-637X/823/1/4}, \href
  {http://adsabs.harvard.edu/abs/2016ApJ...823....4D} {823, 4}

\bibitem[\protect\citeauthoryear{{Dehnen}}{{Dehnen}}{2000}]{wd00}
{Dehnen} W.,  2000, \mn@doi [\aj] {10.1086/301226}, \href
  {http://adsabs.harvard.edu/abs/2000AJ....119..800D} {119, 800}

\bibitem[\protect\citeauthoryear{{Gaia Collaboration} et~al.,}{{Gaia
  Collaboration} et~al.}{2016}]{Gaia+Prusti16}
{Gaia Collaboration} et~al., 2016, \mn@doi [\aap]
  {10.1051/0004-6361/201629272}, \href
  {http://adsabs.harvard.edu/abs/2016A%26A...595A...1G} {595, A1}

\bibitem[\protect\citeauthoryear{{Gaia Collaboration} et~al.,}{{Gaia
  Collaboration} et~al.}{2018b}]{Gaia+Katz18Disc}
{Gaia Collaboration} et~al., 2018b, preprint, \href
  {http://adsabs.harvard.edu/abs/2018arXiv180409380G} {} (\mn@eprint {arXiv}
  {1804.09380})

\bibitem[\protect\citeauthoryear{{Gaia Collaboration}, {Brown}, {Vallenari},
  {Prusti}, {de Bruijne}, {Babusiaux}  \& {Bailer-Jones}}{{Gaia Collaboration}
  et~al.}{2018a}]{Gaia+Brown+18}
{Gaia Collaboration} {Brown} A.~G.~A.,  {Vallenari} A.,  {Prusti} T.,  {de
  Bruijne} J.~H.~J.,  {Babusiaux} C.,   {Bailer-Jones} C.~A.~L.,  2018a,
  preprint, \href {http://adsabs.harvard.edu/abs/2018arXiv180409365G} {}
  (\mn@eprint {arXiv} {1804.09365})

\bibitem[\protect\citeauthoryear{{G{\'o}mez}, {Minchev}, {Villalobos}, {O'Shea}
   \& {Williams}}{{G{\'o}mez} et~al.}{2012}]{Gomez+12}
{G{\'o}mez} F.~A.,  {Minchev} I.,  {Villalobos} {\'A}.,  {O'Shea} B.~W.,
  {Williams} M.~E.~K.,  2012, \mn@doi [\mnras]
  {10.1111/j.1365-2966.2011.19867.x}, \href
  {http://adsabs.harvard.edu/abs/2012MNRAS.419.2163G} {419, 2163}

\bibitem[\protect\citeauthoryear{{G{\'o}mez}, {Minchev}, {O'Shea}, {Beers},
  {Bullock}  \& {Purcell}}{{G{\'o}mez} et~al.}{2013}]{GMO13}
{G{\'o}mez} F.~A.,  {Minchev} I.,  {O'Shea} B.~W.,  {Beers} T.~C.,  {Bullock}
  J.~S.,   {Purcell} C.~W.,  2013, \mn@doi [\mnras] {10.1093/mnras/sts327},
  \href {http://adsabs.harvard.edu/abs/2013MNRAS.429..159G} {429, 159}

\bibitem[\protect\citeauthoryear{{G{\'o}mez}, {White}, {Grand}, {Marinacci},
  {Springel}  \& {Pakmor}}{{G{\'o}mez} et~al.}{2017}]{GWG16}
{G{\'o}mez} F.~A.,  {White} S.~D.~M.,  {Grand} R.~J.~J.,  {Marinacci} F.,
  {Springel} V.,   {Pakmor} R.,  2017, \mn@doi [\mnras]
  {10.1093/mnras/stw2957}, \href
  {http://adsabs.harvard.edu/abs/2017MNRAS.465.3446G} {465, 3446}

\bibitem[\protect\citeauthoryear{{Grand}, {Kawata}  \& {Cropper}}{{Grand}
  et~al.}{2012a}]{gkc12a}
{Grand} R.~J.~J.,  {Kawata} D.,   {Cropper} M.,  2012a, \mn@doi [\mnras]
  {10.1111/j.1365-2966.2012.20411.x}, \href
  {http://adsabs.harvard.edu/abs/2012MNRAS.421.1529G} {421, 1529}

\bibitem[\protect\citeauthoryear{{Grand}, {Kawata}  \& {Cropper}}{{Grand}
  et~al.}{2012b}]{gkc12b}
{Grand} R.~J.~J.,  {Kawata} D.,   {Cropper} M.,  2012b, \mn@doi [\mnras]
  {10.1111/j.1365-2966.2012.21733.x}, \href
  {http://adsabs.harvard.edu/abs/2012MNRAS.426..167G} {426, 167}

\bibitem[\protect\citeauthoryear{{Hart} et~al.,}{{Hart} et~al.}{2017}]{Hart+17}
{Hart} R.~E.,  et~al., 2017, \mn@doi [\mnras] {10.1093/mnras/stx2137}, \href
  {http://adsabs.harvard.edu/abs/2017MNRAS.472.2263H} {472, 2263}

\bibitem[\protect\citeauthoryear{{Hattori}, {Gouda}, {Yano}, {Sakai}, {Tagawa},
  {Baba}  \& {Kumamoto}}{{Hattori} et~al.}{2018}]{Hattori+18}
{Hattori} K.,  {Gouda} N.,  {Yano} T.,  {Sakai} N.,  {Tagawa} H.,  {Baba} J.,
  {Kumamoto} J.,  2018, preprint, \href
  {http://adsabs.harvard.edu/abs/2018arXiv180401920H} {} (\mn@eprint {arXiv}
  {1804.01920})

\bibitem[\protect\citeauthoryear{{Hunt} \& {Bovy}}{{Hunt} \&
  {Bovy}}{2018}]{Hunt+Bovy18}
{Hunt} J.~A.~S.,  {Bovy} J.,  2018, \mn@doi [\mnras] {10.1093/mnras/sty921},
  \href {http://adsabs.harvard.edu/abs/2018MNRAS.477.3945H} {477, 3945}

\bibitem[\protect\citeauthoryear{{Hunt}, {Kawata}, {Grand}, {Minchev},
  {Pasetto}  \& {Cropper}}{{Hunt} et~al.}{2015}]{hkgmpc15}
{Hunt} J.~A.~S.,  {Kawata} D.,  {Grand} R.~J.~J.,  {Minchev} I.,  {Pasetto} S.,
    {Cropper} M.,  2015, \mn@doi [\mnras] {10.1093/mnras/stv765}, \href
  {http://adsabs.harvard.edu/abs/2015MNRAS.450.2132H} {450, 2132}

\bibitem[\protect\citeauthoryear{{Hunt}, {Kawata}, {Monari}, {Grand}, {Famaey}
  \& {Siebert}}{{Hunt} et~al.}{2017}]{Hunt+17}
{Hunt} J.~A.~S.,  {Kawata} D.,  {Monari} G.,  {Grand} R.~J.~J.,  {Famaey} B.,
  {Siebert} A.,  2017, \mn@doi [\mnras] {10.1093/mnrasl/slw257}, \href
  {http://adsabs.harvard.edu/abs/2017MNRAS.467L..21H} {467, L21}

\bibitem[\protect\citeauthoryear{{Hunt}, {Hong}, {Bovy}, {Kawata}  \&
  {Grand}}{{Hunt} et~al.}{2018}]{Hunt+Hong+Bovy+18}
{Hunt} J.~A.~S.,  {Hong} J.,  {Bovy} J.,  {Kawata} D.,   {Grand} R.~J.~J.,
  2018, \mnras submitted

\bibitem[\protect\citeauthoryear{{Katz} et~al.,}{{Katz}
  et~al.}{2018}]{Katz+RV+18}
{Katz} D.,  et~al., 2018, preprint, \href
  {http://adsabs.harvard.edu/abs/2018arXiv180409372K} {} (\mn@eprint {arXiv}
  {1804.09372})

\bibitem[\protect\citeauthoryear{{Kawata}, {Hunt}, {Grand}, {Pasetto}  \&
  {Cropper}}{{Kawata} et~al.}{2014}]{khgpc14}
{Kawata} D.,  {Hunt} J.~A.~S.,  {Grand} R.~J.~J.,  {Pasetto} S.,   {Cropper}
  M.,  2014, \mn@doi [\mnras] {10.1093/mnras/stu1292}, \href
  {http://adsabs.harvard.edu/abs/2014MNRAS.443.2757K} {443, 2757}

\bibitem[\protect\citeauthoryear{{Lindegren} et~al.,}{{Lindegren}
  et~al.}{2018}]{Lindegren+18}
{Lindegren} L.,  et~al., 2018, preprint, \href
  {http://adsabs.harvard.edu/abs/2018arXiv180409366L} {} (\mn@eprint {arXiv}
  {1804.09366})

\bibitem[\protect\citeauthoryear{{Liu}, {Xue}, {Fang}, {van de Ven}, {Wu},
  {Smith}  \& {Carrell}}{{Liu} et~al.}{2012}]{Liu+12}
{Liu} C.,  {Xue} X.,  {Fang} M.,  {van de Ven} G.,  {Wu} Y.,  {Smith} M.~C.,
  {Carrell} K.,  2012, \mn@doi [\apjl] {10.1088/2041-8205/753/1/L24}, \href
  {http://adsabs.harvard.edu/abs/2012ApJ...753L..24L} {753, L24}

\bibitem[\protect\citeauthoryear{{Minchev}, {Quillen}, {Williams}, {Freeman},
  {Nordhaus}, {Siebert}  \& {Bienaym{\'e}}}{{Minchev}
  et~al.}{2009}]{Minchev+09}
{Minchev} I.,  {Quillen} A.~C.,  {Williams} M.,  {Freeman} K.~C.,  {Nordhaus}
  J.,  {Siebert} A.,   {Bienaym{\'e}} O.,  2009, \mn@doi [\mnras]
  {10.1111/j.1745-3933.2009.00661.x}, \href
  {http://adsabs.harvard.edu/abs/2009MNRAS.396L..56M} {396, L56}

\bibitem[\protect\citeauthoryear{{Monari}, {Famaey}, {Siebert}, {Grand},
  {Kawata}  \& {Boily}}{{Monari} et~al.}{2016}]{Monari+16}
{Monari} G.,  {Famaey} B.,  {Siebert} A.,  {Grand} R.~J.~J.,  {Kawata} D.,
  {Boily} C.,  2016, \mn@doi [\mnras] {10.1093/mnras/stw1564}, \href
  {http://adsabs.harvard.edu/abs/2016MNRAS.461.3835M} {461, 3835}

\bibitem[\protect\citeauthoryear{{Monari}, {Kawata}, {Hunt}  \&
  {Famaey}}{{Monari} et~al.}{2017}]{Monari+17}
{Monari} G.,  {Kawata} D.,  {Hunt} J.~A.~S.,   {Famaey} B.,  2017, \mn@doi
  [\mnras] {10.1093/mnrasl/slw238}, \href
  {http://adsabs.harvard.edu/abs/2017MNRAS.466L.113M} {466, L113}

\bibitem[\protect\citeauthoryear{{Quillen} et~al.,}{{Quillen}
  et~al.}{2018}]{Quillen+18}
{Quillen} A.~C.,  et~al., 2018, preprint, \href
  {http://adsabs.harvard.edu/abs/2018arXiv180510236Q} {} (\mn@eprint {arXiv}
  {1805.10236})

\bibitem[\protect\citeauthoryear{{Ramos}, {Antoja}  \& {Figueras}}{{Ramos}
  et~al.}{2018}]{Ramos+18}
{Ramos} P.,  {Antoja} T.,   {Figueras} F.,  2018, preprint, \href
  {http://adsabs.harvard.edu/abs/2018arXiv180509790R} {} (\mn@eprint {arXiv}
  {1805.09790})

\bibitem[\protect\citeauthoryear{{Reid} et~al.,}{{Reid} et~al.}{2014}]{rmbzd14}
{Reid} M.~J.,  et~al., 2014, \mn@doi [\apj] {10.1088/0004-637X/783/2/130},
  \href {http://adsabs.harvard.edu/abs/2014ApJ...783..130R} {783, 130}

\bibitem[\protect\citeauthoryear{{Sartoretti} et~al.,}{{Sartoretti}
  et~al.}{2018}]{Sartoretti+18}
{Sartoretti} P.,  et~al., 2018, preprint, \href
  {http://adsabs.harvard.edu/abs/2018arXiv180409371S} {} (\mn@eprint {arXiv}
  {1804.09371})

\bibitem[\protect\citeauthoryear{{Sch{\"o}nrich} \& {Dehnen}}{{Sch{\"o}nrich}
  \& {Dehnen}}{2017}]{Schoenrich+Dehnen18}
{Sch{\"o}nrich} R.,  {Dehnen} W.,  2017, preprint, \href
  {http://adsabs.harvard.edu/abs/2017arXiv171206616S} {} (\mn@eprint {arXiv}
  {1712.06616})

\bibitem[\protect\citeauthoryear{{Sharma}, {Bland-Hawthorn}, {Johnston}  \&
  {Binney}}{{Sharma} et~al.}{2011}]{sbhjb11}
{Sharma} S.,  {Bland-Hawthorn} J.,  {Johnston} K.~V.,   {Binney} J.,  2011,
  \mn@doi [\apj] {10.1088/0004-637X/730/1/3}, \href
  {http://adsabs.harvard.edu/abs/2011ApJ...730....3S} {730, 3}

\bibitem[\protect\citeauthoryear{{Tian} et~al.,}{{Tian} et~al.}{2017}]{Tian+17}
{Tian} H.-J.,  et~al., 2017, \mn@doi [Research in Astronomy and Astrophysics]
  {10.1088/1674-4527/17/11/114}, \href
  {http://adsabs.harvard.edu/abs/2017RAA....17..114T} {17, 114}

\bibitem[\protect\citeauthoryear{{Trick}, {Coronado}  \& {Rix}}{{Trick}
  et~al.}{2018}]{Trick+18}
{Trick} W.~H.,  {Coronado} J.,   {Rix} H.-W.,  2018, preprint, \href
  {http://adsabs.harvard.edu/abs/2018arXiv180503653T} {} (\mn@eprint {arXiv}
  {1805.03653})

\bibitem[\protect\citeauthoryear{{Wada}, {Baba}  \& {Saitoh}}{{Wada}
  et~al.}{2011}]{wbs11}
{Wada} K.,  {Baba} J.,   {Saitoh} T.~R.,  2011, \mn@doi [\apj]
  {10.1088/0004-637X/735/1/1}, \href
  {http://adsabs.harvard.edu/abs/2011ApJ...735....1W} {735, 1}

\bibitem[\protect\citeauthoryear{{Wang}, {L{\'o}pez-Corredoira}, {Carlin}  \&
  {Deng}}{{Wang} et~al.}{2018}]{Wang+18}
{Wang} H.,  {L{\'o}pez-Corredoira} M.,  {Carlin} J.~L.,   {Deng} L.,  2018,
  \mn@doi [\mnras] {10.1093/mnras/sty739}, \href
  {http://adsabs.harvard.edu/abs/2018MNRAS.tmp..721W} {}

\bibitem[\protect\citeauthoryear{{Widrow}, {Gardner}, {Yanny}, {Dodelson}  \&
  {Chen}}{{Widrow} et~al.}{2012}]{Widrow+12}
{Widrow} L.~M.,  {Gardner} S.,  {Yanny} B.,  {Dodelson} S.,   {Chen} H.-Y.,
  2012, \mn@doi [\apjl] {10.1088/2041-8205/750/2/L41}, \href
  {http://adsabs.harvard.edu/abs/2012ApJ...750L..41W} {750, L41}

\bibitem[\protect\citeauthoryear{{Williams} et~al.,}{{Williams}
  et~al.}{2013}]{Williams+13}
{Williams} M.~E.~K.,  et~al., 2013, \mn@doi [\mnras] {10.1093/mnras/stt1522},
  \href {http://adsabs.harvard.edu/abs/2013MNRAS.436..101W} {436, 101}

\bibitem[\protect\citeauthoryear{{Xu}, {Newberg}, {Carlin}, {Liu}, {Deng},
  {Li}, {Sch{\"o}nrich}  \& {Yanny}}{{Xu} et~al.}{2015}]{XNC15}
{Xu} Y.,  {Newberg} H.~J.,  {Carlin} J.~L.,  {Liu} C.,  {Deng} L.,  {Li} J.,
  {Sch{\"o}nrich} R.,   {Yanny} B.,  2015, \mn@doi [\apj]
  {10.1088/0004-637X/801/2/105}, \href
  {http://adsabs.harvard.edu/abs/2015ApJ...801..105X} {801, 105}

\bibitem[\protect\citeauthoryear{{de la Vega}, {Quillen}, {Carlin},
  {Chakrabarti}  \& {D'Onghia}}{{de la Vega} et~al.}{2015}]{delaVega+15}
{de la Vega} A.,  {Quillen} A.~C.,  {Carlin} J.~L.,  {Chakrabarti} S.,
  {D'Onghia} E.,  2015, \mn@doi [\mnras] {10.1093/mnras/stv2055}, \href
  {http://adsabs.harvard.edu/abs/2015MNRAS.454..933D} {454, 933}

\makeatother
\end{thebibliography}







\bsp	
\label{lastpage}
\end{document}